\documentclass[conference]{IEEEtran}
\IEEEoverridecommandlockouts
\usepackage{url}
\usepackage{hyperref}
\usepackage{cite}
\usepackage{amsmath,amssymb,amsfonts,mathrsfs}
\usepackage{mathtools}
\usepackage{algorithmic}
\usepackage{float}
\usepackage{tikz}
\usepackage{multirow}
\usetikzlibrary{matrix,calc}
\usepackage[export]{adjustbox}
\usepackage{graphicx}
\usepackage{float}
\usepackage{textcomp}
\usepackage{xcolor}
\def\BibTeX{{\rm B\kern-.05em{\sc i\kern-.025em b}\kern-.08em
    T\kern-.1667em\lower.7ex\hbox{E}\kern-.125emX}}

\begin{document}

\title{Hot-Starting the Ac Power Flow with Convolutional Neural Networks}

\author{\IEEEauthorblockN{Liangjie Chen, Joseph Euzebe Tate}
}

\maketitle

\begin{abstract}
Obtaining good initial conditions to solve the Newton-Raphson (NR) based ac power flow (ACPF) problem can be a very difficult task. In this paper, we propose a framework to obtain the initial bus voltage magnitude and phase values that decrease the solution iterations and time for the NR based ACPF model, using the dc power flow (DCPF) results and one dimensional convolutional neural networks (1D CNNs).
We generate the dataset used to train the 1D CNNs by sampling from a distribution of load demands, and by computing the DCPF and ACPF results for each sample. Experiments on the IEEE 118-bus and \textsc{Pegase} 2869-bus study systems show that we can achieve 33.56\% and 30.06\% reduction in solution time, and 66.47\% and 49.52\% reduction in solution iterations per case, respectively. We include the 1D CNN architectures and the hyperparameters used, which can be expanded on by the future studies on this topic.
\\*
\end{abstract}


\section{Introduction}
\IEEEPARstart{I}{n} power flow studies, the linearized dc power flow (DCPF) model offers compelling computational advantages over the nonlinear ac power flow (ACPF) model based on the Newton-Raphson (NR) method, which requires iterative solutions. However, the DCPF model results become more inaccurate in the cases where its assumptions no longer hold true, e.g. with high \textit{R/X} ratios, large phase angles and heavy or light loads present \cite{dcpf}. Additionally, it can be difficult to find the initial conditions for NR based ACPF to converge \cite{nr}. This paper presents a framework that produces initial conditions that reduce the ACPF iterations and solution times, and is generalizable to grid topologies of different sizes. We use feed-forward artificial neural networks, specifically, one dimensional convolutional neural networks (1D CNNs) to achieve these goals.

Feed-forward neural networks with nonlinear activation functions can be used to approximate any continuous functions \cite{cnn}. CNNs, in particular, can capture the local features of interest more effectively than other feed-forward neural networks such as Multilayer Perceptrons (MLPs), as proven in many computer vision applications \cite{imagenet}. In the context of this paper, an example of such a local feature is the small voltage angle difference between neighboring buses. Additionally, we choose the 1D CNNs since the signals in the buses we are interested in (real and reactive power, voltage magnitude and phase) can be represented as vectors.    

For 1D CNNs trained on DCPF results as input data and corresponding ACPF results as ground truth values, our goal is to produce bus voltage values that, when used as initial conditions to run NR ACPF, result in lower solution iterations and time compared to cold-start (also known as ``flat start") conditions, i.e., $1.0\angle 0.0^{\circ}$ for all load (PQ) bus voltages \cite{overbye}, or warm-start conditions such as the ones generated by DCPF or past solutions. 

Our proposed method considers only the fluctuations of PQ bus demands, i.e., we vary the real and reactive power demand levels at each load bus and solve for the voltage magnitude and phase at each bus, for a specific set of load bus demands. Fluctuations in the generator (PV) buses, e.g. real power injection variations from wind generation or changes in bus voltage magnitudes are not included in our data, but can be incorporated relatively easily in future studies. In Section \ref{s2}, we review some related studies and provide a high-level description of the proposed model. We then present the data generation process, CNN training and the hot-start ACPF procedure in detail in Sections \ref{s3} and \ref{s4}. Finally, in Section \ref{s5}, we present some results based on the IEEE 118-bus and the \textsc{Pegase} 2869-bus systems \cite{pegase1, pegase2} available from \textsc{Matpower} \cite{matpower}. We conclude by giving a short summary and pointing out some limitations of our proposed method, as well as potential directions for future studies on this topic. \\

\section{Background and Proposed Method}\label{s2}

\subsection{Related Work}\label{s2-a}
There have been attempts at either improving the DCPF results or directly predicting ACPF results using artificial neural networks, specifically MLPs, in the past \cite{annpf, saudi, small}. However, they suffer from one or more of the following deficiencies: insufficient dataset size, poorly justified MLP input feature selection which could potentially lead to numerical instability during training, arbitrary and/or unclear performance criteria, and small system sizes where full ACPF can be easily and efficiently computed. Aside from these, there have also been studies on solving or reducing the optimal power flow problem using artificial neural networks \cite{opf-gnn, opf-meta, deepOpf}.

Compared to the MLPs that we trained on the same dataset (formatted differently from as shown in Section \ref{s3-c}, for implementation purposes), 1D CNNs are capable of producing results with $\Delta\mathcal{L}$
(see Equation \ref{delL}) that are almost 10 times smaller than those produced by MLPs. Generally speaking, although 1D CNNs take longer to train, they often have fewer model parameters, thus, lower memory requirement than MLPs to achieve the same or better results. This fact can be a major advantage for extremely large systems.

\subsection{Proposed Method}\label{s2-b}
\begin{figure}[H]
    \centering
    \includegraphics[trim={0cm 1cm 17cm .75cm},clip,scale=0.575,center]{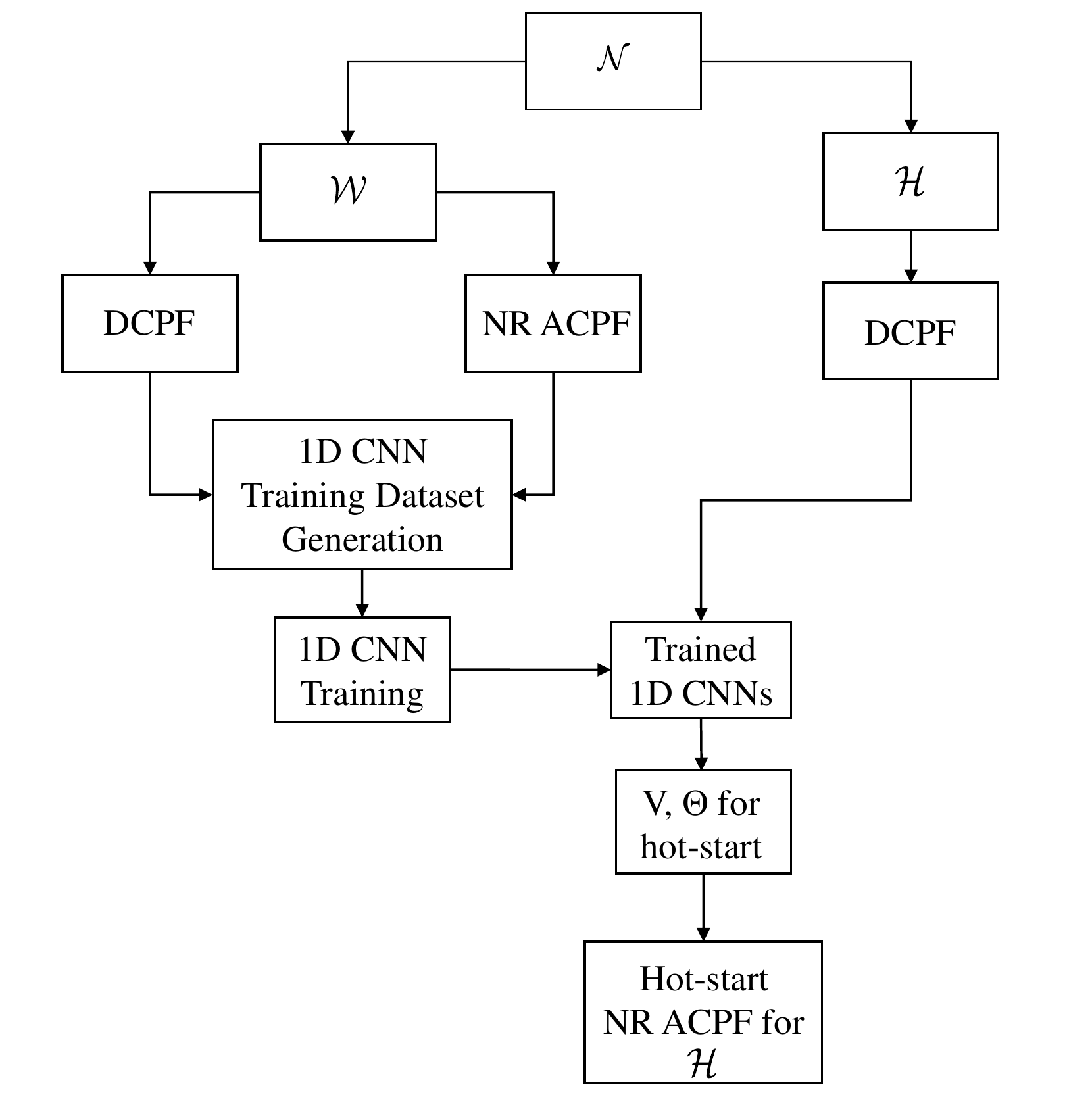}
    \caption{Proposed method}
    \label{flowchart}
\end{figure}

Our proposed method is as follows. Suppose that for a specific system with $L$ buses, we have $N$ different load conditions which can be solved by warm-start ACPF. We need to find the respective load bus voltage magnitude and phase values that meet the mismatch tolerance for these load conditions. Let $\mathcal{N}$ represent this set of $N$ load conditions. First, we take a subset of $\mathcal{N}$, denoted $\mathcal{W}$ for ``warm-start," and let the remaining $\mathcal{N \setminus W}$ be $\mathcal{H}$ for ``hot-start." Let $T = \vert \mathcal{W} \vert$, i.e., the number of load conditions in $\mathcal{W}$. Next, we compute the DCPF results for all load conditions in $\mathcal{N}$, and compute ACPF results with DCPF results as initial conditions (i.e., warm-start) for all load conditions in $\mathcal{W}$. We then use the DCPF and ACPF results corresponding to the load conditions in $\mathcal{W}$, as input data and output targets to train the 1D CNNs. Finally, for load conditions in $\mathcal{H}$, for which we only have the DCPF results, we produce the hot-start conditions for them by passing their DCPF results and corresponding load conditions into the trained 1D CNNs, and compute the ACPF results for these load conditions with the hot-start conditions. This process is shown in Figure \ref{flowchart}. Once the 1D CNNs are trained, any new load conditions from the same system but are not in $\mathcal{N}$,  would follow the path that the load conditions in $\mathcal{H}$ take in Figure \ref{flowchart} --- first compute the DCPF results, then use the trained 1D CNNs to generate the initial conditions to compute the ACPF results. 

The reason to treat the CNN-predicted values as hot-start conditions for ACPF instead of as finished products is so that we have a fair comparison between our method and the ACPF model with warm-start conditions, since both, if they converge successfully, will have a final mismatch within the same tolerance level. Clearly, since ACPF is computationally expensive, and neural network training with more data takes more time (empirically, quasi-linearly on a single GPU), we would like to find the minimum $T$ that provides reasonably good hot-start conditions for load conditions in $\mathcal{H}$ on the chosen CNN architecture. We also point out that in this study, we are investigating the feasibility of this approach by evaluating its effectiveness on the IEEE 118-bus and, more realistically, the larger \textsc{Pegase} 2869-bus systems; we are not focused on devising the best possible CNN architecture, which will be discussed in Section \ref{s4}. 
\\

\section{Data Generation}\label{s3}

\subsection{Load Fluctuation Modeling}\label{s3-a}
As discussed in Section \ref{s2-b}, the first step to create our dataset is to generate load demand fluctuations for all PQ buses in the system. For the $i$-th PQ bus in a \textsc{Matpower} case, we extract the default real power demand value, $P_i$ as mean real power demand, and compute the corresponding standard deviation $\sigma_{i}$ as follows (from \cite{pqvar}):
\begin{equation}\label{eq1}
\sigma_{i} = 5.44130 + 0.17459\sqrt{|P_i|}+0.001673{|P_i|}
\end{equation}

Next, we generate $N$ samples from the Gaussian distribution $\mathcal{N}(P_i,\, \sigma^2_{i})$ to create the real power demand fluctuations $[P^{(1)}_i,\, P^{(2)}_i,\, ...,\, P^{(N)}_i]$ for the $i$-th bus if it is a PQ bus; otherwise we keep the default value, i.e., $P^{(k)}_i = P_i$ for $1 \leq k \leq N$. We generate the demand realizations for each bus independently and with a fixed random seed for reproducibility. We can now represent our real power demand fluctuations for all $L$ buses in a given system as a matrix:
\[
\mathbf{P} = \begin{bmatrix}
P^{(1)}_1 & \dots & P^{(N)}_1 \\
\vdots & \ddots & \vdots \\
P^{(1)}_L & \dots & P^{(N)}_L
\end{bmatrix}
\]

To solve the power flow problem, the known values for each load bus are real power P and reactive power Q. Therefore, we now need to generate the \textbf{Q} matrix to similarly represent the fluctuations in reactive power. First, we generate p.f., a vector containing $N$ samples of lagging power factor, with a Gaussian distribution $\mathcal{N}(\mu=1.0,\, \sigma=0.05)$ truncated between $[0.7, 1.0]$. The choices of $\mu$ and $\sigma$ are based on the distributions of power factor values for all PQ buses in the \textsc{Matpower} cases. We choose the truncation lower bound of 0.7 because a utility would step in and fix the power factors lower than that (e.g., by penalizing businesses to discourage low power factors) to avoid loss \cite{pf, lowpf}. We then calculate each entry of the \textbf{Q} matrix as follows:
\begin{equation}\label{eq2}
Q^{(k)}_i = P^{(k)}_i \cdot \tan (\arccos(\text{p.f.}^{(k)})) \\
\end{equation}

Similar to the $\mathbf{P}$ matrix, we keep the default value for $Q^{(k)}_i$, $1 \leq k \leq N$ if the $i$-th bus is not a PQ bus.

\subsection{Power Flow Computation}\label{s3-b}
As discussed in Section \ref{s2}, we first run DCPFs based on the P, Q values in $\mathcal{N}$ and warm-start ACPF for $\mathcal{W}$, then use the DCPF and ACPF results corresponding to P, Q values in $\mathcal{W}$ to train a CNN. For bench-marking purposes, we also run warm-start ACPF for P, Q values in $\mathcal{H}$ and prensent the warm-start performance in Section \ref{s5}. Once a model is trained, we can then follow Section \ref{s2-b} and only run hot-start ACPF for the load levels in $\mathcal{H}$ or any new load conditions. 

We use \textsc{Matpower}'s \texttt{rundcpf} and \texttt{runpf} functions to perform the dc and ac power flow computations with the mismatch tolerance for \texttt{runpf} set at $10^{-3}$ per unit. For the $k$-th execution of \texttt{rundcpf} and \texttt{runpf}, we replace the default $P_i$, $Q_i$ values of the $i$-th bus with $P^{(k)}_i$ and $Q^{(k)}_i$ from the $k$-th column of $\mathbf{P}$ and \textbf{Q}. We do not change any other values. We then collect the solved voltage magnitude and phase values --- ${V_{DC}}^{(k)}_{i} $ (which will always be 1.0) and ${\Theta_{DC}}^{(k)}_{i}$ (in radians) from the DCPF results, along with ${V_{AC}}^{(k)}_{i}$ and ${\Theta_{AC}}^{(k)}_{i}$ (in radians) from the ACPF results. The reason to extract voltage phase values in radians is that large negative phase values in degrees will easily cause exploding loss when passed through the ELU activation function (Equation \ref{ELU}). We collect the DCPF solution time, $t_{DC}$ as a vector of length $T$ to calculate the average hot-start ACPF time as described in Section \ref{s5}. We also collect the ACPF solution time and iterations corresponding to the load levels in $\mathcal{H}$, $t_{AC, warm}$, $n_{AC, warm}$, as vectors with length $(N-T)$, which are used to compare with the hot-start ACPF performances. 

Finally, since ${\Theta_{DC}}^{(k)}_{i}$, ${\Theta_{AC}}^{(k)}_{i}$ are in radians, we would like to ensure all the input data are in a similar range, so that the model parameter updates are input unit agnostic. Therefore, we perform the following data processing steps. We subtract 1.0, the nominal voltage from all ${V_{DC}}^{(k)}_{i}$ and ${V_{AC}}^{(k)}_{i} $, so that they have near 0 mean. We do not normalize the input and target voltage magnitude and phase values to be within a certain range (e.g., in computer vision applications, we could normalize the input data values to be in $[0,1]$) since the lower and upper bound of the ground truth, which are required for the normalization are not known \textit{a priori}. We also compute the $\mathbf{P_d}$ and $\mathbf{Q_d}$ matrices with entries ${P_d}^{(k)}_{i} = P^{(k)}_i - P_i$ and ${Q_d}^{(k)}_{i} = Q^{(k)}_i - Q_i$ in per unit, respectively. We construct the following matrices with the same dimensions as $\mathbf{P}$ and \textbf{Q}, for $1 \leq i \leq L$, and $1 \leq k \leq N$: 
\[
\mathbf{V_{DC}} = \begin{bmatrix}
{V_{DC}}^{(k)}_{i} \\
\end{bmatrix}
\hspace{4em}
\mathbf{\Theta_{DC}} = \begin{bmatrix}
{\Theta_{DC}}^{(k)}_{i}
\end{bmatrix}
\]
\[
\mathbf{V_{AC}} = \begin{bmatrix}
{V_{AC}}^{(k)}_{i} \\
\end{bmatrix}
\hspace{4em}
\mathbf{\Theta_{AC}} = \begin{bmatrix}
{\Theta_{AC}}^{(k)}_{i}
\end{bmatrix}
\]

Note that in Section \ref{s2-b}, we assumed none of the load conditions in $\mathcal{N}$ causes the non-convergence of ACPF. Therefore, $\mathbf{V_{AC}}, \mathbf{\Theta_{AC}}$ contain the same number of samples, $N$, as $\mathbf{V_{DC}}, \mathbf{\Theta_{DC}}$. However, this assumption is not realistic, since not all load conditions are guaranteed to converge with warm-start conditions. Therefore, if the $k$-th ACPF execution fails to converge and it is a load condition in $\mathcal{W}$, we add it to a set $\mathcal{F}$ that contains all load conditions that fail to converge. We stop our data generation process once $N$ samples successfully converges. If $\mathcal{F} \neq \emptyset$, we also note the successful ACPF convergence rate.

For the bus voltage magnitudes and phase values generated by the 1D CNNs, we shift the voltage magnitude values back up by 1.0 and convert the phase values back to be in degrees as required by \textsc{Matpower}, before performing the ACPF computations with these as the initial conditions. 

\subsection{Dataset Format}\label{s3-c}
\begin{figure}[h]
    \centering
    \begin{tikzpicture}[every node/.style={anchor=north east,fill=white,minimum width=4em,minimum height=3em}]
    \matrix (mA) [draw,matrix of math nodes]
    {\mathbf{Q_{d}} \\};
    \matrix (mB) [draw,matrix of math nodes] at ($(mA.south west)+(0.15, 0.6)$)
    {\mathbf{P_{d}}\\};
    \matrix (mC) [draw,matrix of math nodes] at ($(mB.south west)+(0.15, 0.6)$)
    {\mathbf{\Theta_{DC}} \\};
    \matrix (mD) [draw,matrix of math nodes]  at ($(mC.south west)+(0.15, 0.6)$)
    {\mathbf{V_{DC}}\\};
    
    \draw[dashed](mA.north east)--(mD.north east);
    \draw[dashed](mA.north west)--(mD.north west);
    \draw[dashed](mA.south east)--(mD.south east);
    \end{tikzpicture} \\
    \caption{Dataset $\mathbf{X}$}
    \label{X}
\end{figure}
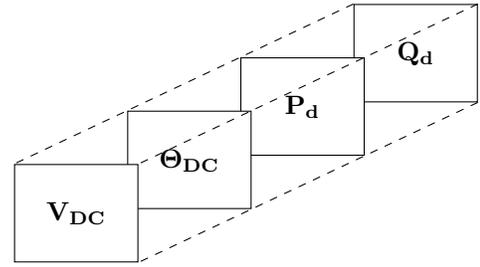

\begin{figure}[h]
    \centering
    \includegraphics[trim={2cm 9cm 24cm 3.2cm},clip,scale=1,center]{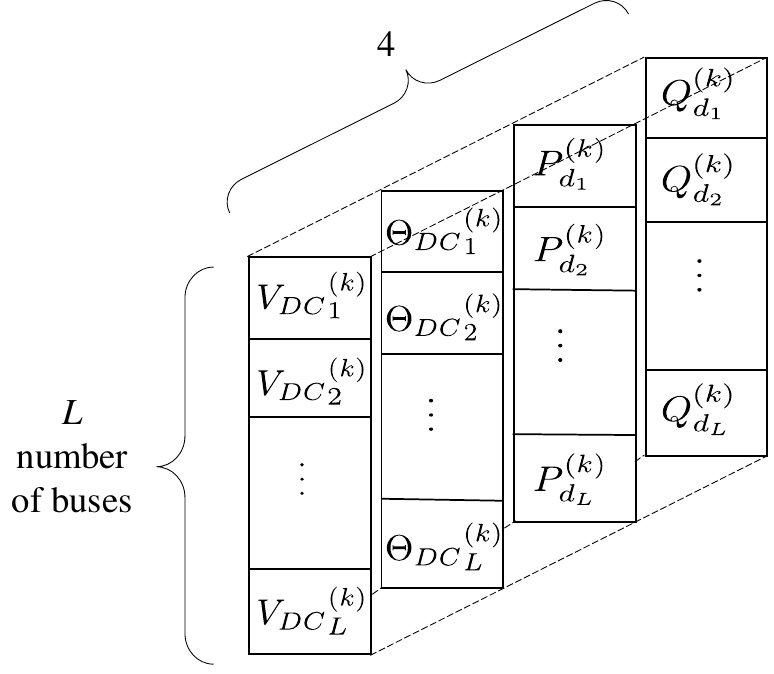}
    \caption{The $k$-th sample of the dataset $\mathbf{X}$}
    \label{Xk}
\end{figure}

We now form the dataset for training. The input to the 1D CNNs, $\mathbf{X}$, containing the offset DCPF bus voltage values, $\mathbf{P_d}$ and $\mathbf{Q_d}$ is a tensor with dimension $L \times 4 \times N$ and the format shown in Figure \ref{X}. Specifically, the $k$-th sample in $\mathbf{X}$, $1 \leq k \leq N$, is shown in Figure \ref{Xk}. 

We train two 1D CNNs with the identical architecture (shown in Figure \ref{model}), both with $\mathbf{X}$ as the input --- one for producing hot-start bus voltage magnitudes (V model) and the other for bus voltage phases ($\Theta$ model). This approach is to ensure that the 1D CNN model parameter updates are computed based on the loss with respect to distinct targets, $\mathbf{V_{AC}}$ and $\mathbf{\Theta_{AC}}$, even though we offset $\mathbf{V_{AC}}$ to have near 0 mean.

Additionally, we reshape $\mathbf{X}$ to add a ``width" dimension of length 1, in order to fit the  Width -- Height -- Channel -- Samples format, which the machine learning library we used, Flux \cite{flux}, requires. \\

\section{CNN Training} \label{s4}

\subsection{Loss Function and Performance Criteria} \label{s4-a}
The loss function $\ell$ we use to train our model is the squared $\mathcal{L}_2$ norm of the difference between predicted values and ground truth: $\ell(\hat{y_i} , y_i) = \Vert \hat{y_i} - y_i \Vert^2_2$, where $\hat{y_i}$ is the prediction and $y_i$ is the ground truth for the $i$-th element. This loss is commonly used for optimization in regression problems, and we found that it outperforms the mean square error, another popular loss function in regression problems, in terms of $\Delta\mathcal{L}$ defined below. We do not have an accuracy measure since there is no robust and generalizable way of establishing such a criteria in our problem. The downside of using relative measurements (e.g. mean absolute relative error), in particular in this problem, is that the error would explode when the denominator is close or equal to 0, which is not uncommon in the voltage angle or offset voltage magnitude values. Instead, we compare the initial and final $\mathcal{L}_2$ norms on the test set to see how much the norm decreased at the end of training by $\Delta\mathcal{L}$ in Equation \ref{delL}, where $\mathcal{L}_i$ is the $\mathcal{L}_2$ norm between dc and ac power flow results for $\mathcal{H}$, and $\mathcal{L}_f$ is the $\mathcal{L}_2$ norm between the predicted hot-start conditions and true ac power flow results for $\mathcal{H}$ (both $\mathcal{L}_i$ and $\mathcal{L}_f$ are the average of voltage magnitude and phase results), which we computed for benchmarking purposes but will not be available in practice.
\begin{equation}\label{delL}
\Delta \mathcal{L} = \frac{\mathcal{L}_f}{\mathcal{L}_i} \times 100\%
\end{equation}

The effectiveness of the 1D CNNs, however, is best demonstrated by the performance of computing ACPF results for $\mathcal{H}$, i.e., by how much the 1D CNN produced bus voltage values can decrease the ACPF iterations and solution time (as discussed in Section \ref{s5}).

\subsection{CNN Model Architecture and Hyperparameter Selection} \label{s4-b}
We determine the 1D CNN model architecture and select the hyperparameters in the following way. (To see the meaning of the terms used here, please refer to the Appendix for a brief overview of CNNs.) For the IEEE 118-bus system, we first use a training set with $T=2000$, and a set of hyperparameters commonly seen in machine learning applications \cite{batchsizes, Goodfellow2015DL}: initial learning rate $\eta = 10^{-3}$, batch size of 64, and maximum epochs of 500. With these hyperparameters selected, we train multiple 1D CNNs with different combinations of number of convolutional layers, kernel sizes and number of channels. We then make a decision on these hyperparameters based on the validation set losses of each candidate architecture. We repeat this process for the \textsc{Pegase} 2869-bus system. Once we arrive at a satisfactory final validation loss, we go back to test our initial hyperparameter choices, similarly by the validation loss. In our case studies, we keep the initial learning rate $\eta$ and maximum epoch the same as initially chosen, and only change the batch size from 64 to 32.

\begin{figure}[h]
    \includegraphics[trim={0cm 1cm 0cm .5cm},clip,scale=0.45,center]{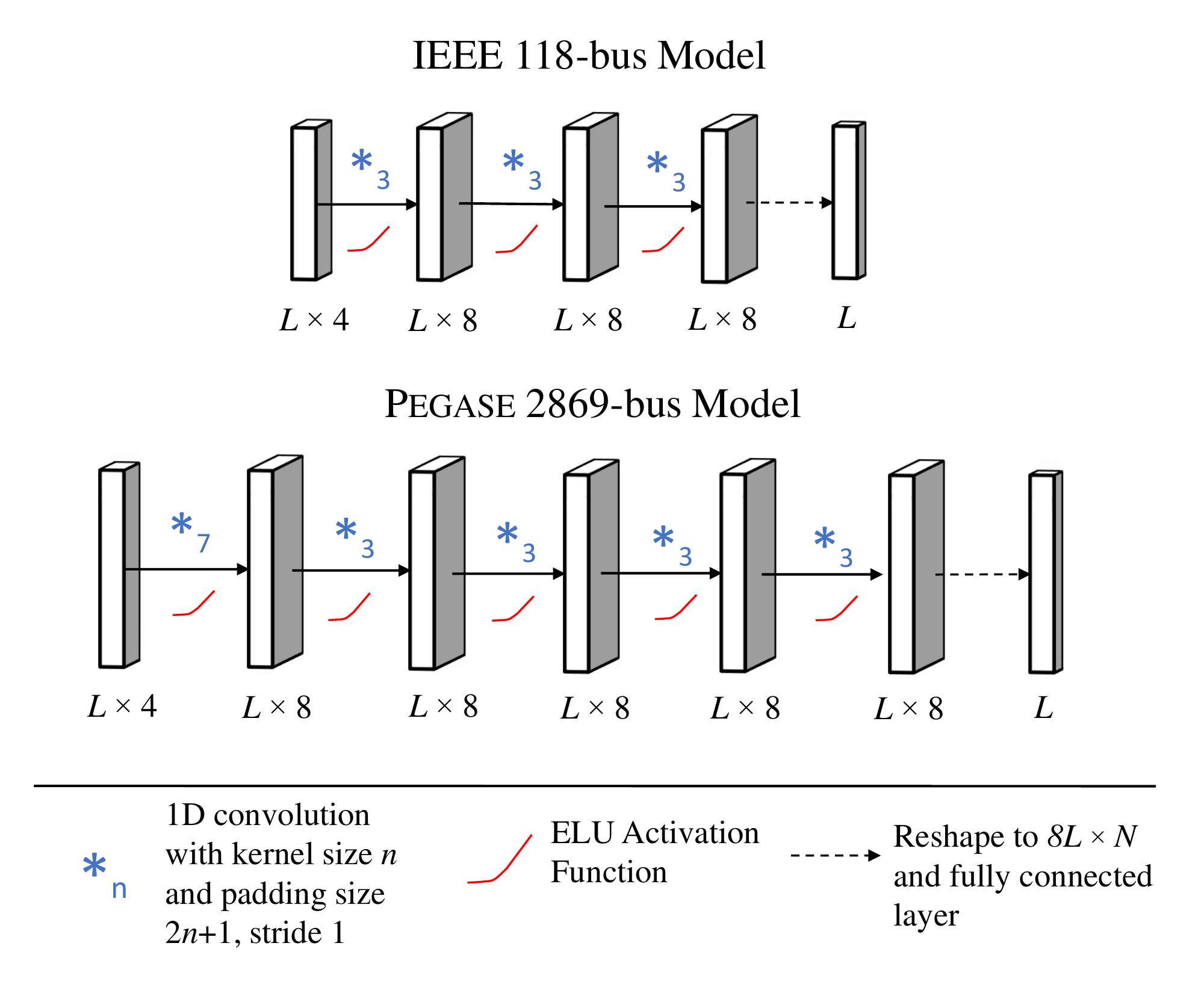}
    \caption{1D CNN architectures. We train two identical models (one trained with $\mathbf{V}_{AC}$ as output target the other with $\mathbf{\Theta}_{AC}$ as target) for each case, and only one is shown here}
    \label{model}
\end{figure}

We will now discuss our chosen 1D CNN architectures in detail. Figure \ref{model} shows the \textit{Height} (which equals to the number of buses, $L$) and \textit{Channel} dimensions in each convolutional layer, and a single sample (as in Figure \ref{Xk}) as the input to each 1D CNN model. Each of the outputs in Figure \ref{model} is a vector containing the predicted voltage or phase values. Thus, the output would be a matrix when we feed more than one input samples into the 1D CNN. For the \textsc{Pegase} 2869-bus system, we use a deeper architecture with 5 convolutional layers and a final fully connected layer. The first convolutional layer has kernel size 7 with channel size 8, zero padding size 3 and stride 1, and identical remaining convolutional layers with kernel size 3, channel size 8, zero padding size 1 and stride 1. The CNN model for the smaller IEEE 118-bus system has three identical convolutional layers with kernel size 3, zero padding size 1 and stride 1, i.e., the same as the second to the fifth convolutional layers of the larger model. The fully connected layer for both architectures first reshapes the data to a vector of length $8L$ (for 1 sample), then produces the final bus voltage magnitude and phase values. Empirically, adding more than 8 channels to the convolutional layers result in worse $\Delta\mathcal{L}$ of the validation set. We apply the zero paddings and stride 1 to all convolutional layers, and omit pooling layers since we want to keep the hidden layer dimension the same as the input feature vector dimension ($L$, i.e., number of buses), throughout the architecture. The reason for this choice is that CNNs for the systems with odd number of buses, after convolution, pooling and up-sampling operations with strides larger than 1, will produce predictions with one extra or one fewer value, i.e., an extra or a missing bus. This is not to say that such architectures will never work in our case, since we can, for example, apply asymmetric padding to solve this off-by-1 caveat. However, operations such as pooling and upsampling, even if they improve the final predictions, will make the justification of our method more difficult. In particular, the pooling operation is lossy, since it downsamples the input data into a low-dimensional representation.

Since the number of positive and negative values in our datasets are roughly equal, the nonlinear activation function $f$ we use has to account for both. We compared the performance and rate of convergence of multiple popular activation functions --- Rectified Linear Unit (ReLU) \cite{relu}, LeakyReLU \cite{leakyrelu}, and Exponential Linear Unit (ELU) \cite{elu}, and chose ELU with the default parameter $\alpha=1.0$, which has the piecewise definition and derivative in Equations \ref{ELU} and \ref{dELU}, respectively.
\begin{equation}\label{ELU}
f(x) = \left\{
\begin{array}{cl}
x & \text{if } x > 0,\\
\alpha (e^x - 1) & \text{if } x \leq 0
\end{array} \right.
\end{equation} 
\begin{equation}\label{dELU}
f'(x) = \left\{
\begin{array}{cl}
1 & \text{if } x > 0,\\
\alpha e^x & \text{if } x \leq 0
\end{array} \right.
\end{equation} 

\begin{figure}[H]
    \includegraphics[trim={2cm 0.5cm 2cm 1cm},clip,scale=0.4,center]{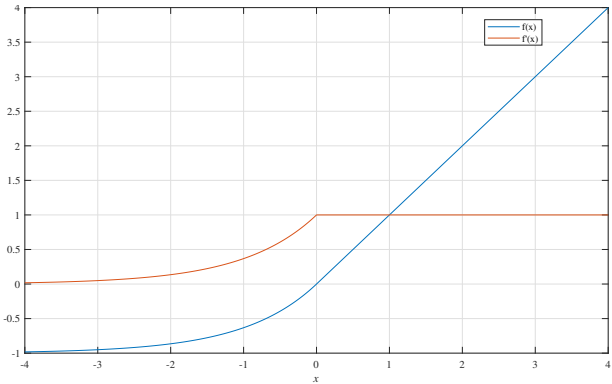}
    \caption{Exponential Linear Unit ($\alpha=1.0$) and its derivative}
    \label{case118_va}
\end{figure}

With the CNN architecture fixed for a particular system, the most important hyperparameter left in this problem is the size of $T$, i.e., we want to find the optimal trade-off point between training set size (with which the training time scales quasi-linearly, as Tables \ref{table-118} and \ref{table-2869} in the next section shows) and the quality of the trained model (i.e., how much its prediction can decrease solution time and iterations). Since the training and validation targets are the ACPF results, a smaller $T$ value means a smaller number of warm-start ACPF we have to run in data generation process, as mentioned in Section \ref{s3}. 

Since we also computed the ACPF results for $\mathcal{H}$ for benchmarking purposes, we have the ground truth values for all $N$ samples. Thus, we can use the true ACPF results of the load conditions in $\mathcal{H}$ as the test set (in practice, we would not have these data since we do not compute warm-start ACPF for all samples). 
We separate $\mathcal{W}$ into training and validation sets, with a 90/10 split, i.e., the training set will have $0.9T$ number of samples and the validation set will have $0.1T$ number of samples. Since the dataset is generated with Gaussian distributions instead of gathered from real systems, the training and validation loss values throughout the training are extremely close, and as a result, we do not need a large validation set to adjust model hyperparameters, or use methods such as cross-validation during training. 

Since learning hot-start conditions is a rather uncommon application of CNNs, we cannot take advantage of previously trained models and use transfer learning \cite{transferlearning1, transferlearning2} to accelerate training. Therefore, we initialize the CNN parameters with the Xavier Initialization \cite{xavier}, and train the models from scratch with the Adam optimizer \cite{adam}. During training, we randomly shuffle the batch indices with a fixed random seed before each epoch starts. We also use the following learning rate decay policy \cite{alexnet}: if training set loss does not decrease for 5 consecutive epochs, we decrease the learning rate $\eta$ by a factor of 10 (until $\eta = 10^{-9}$) to encourage the model to jump out of local minimums. Training is terminated if the elapsed epochs reach maximum of 500 or $\Delta \mathcal{L}$ of the validation set becomes less than 0.01\%. 
\\

\section{Case Study} \label{s5}
The dataset generation and power flow computations are done with MATLAB and \textsc{Matpower} on a local computer with Intel Core i7-8750H CPU and 32 GB RAM. The 1D CNN training is done on a Compute Canada cluster \cite{graham} using a single Intel Xeon Gold 5120 CPU and a single NVIDIA Tesla V100 GPU with 16 GB of GPU memory. We use Flux \cite{flux}, an open-source machine learning library developed in the Julia language \cite{julia}, for the 1D CNN implementation, training and inference. 

The following tables show the average solution time and average iteration count for $N=10000$ samples, all of which successfully converged. The first row contains the warm-start performance, where $t_{avg} = \Bar{t}_{AC,warm}$, i.e., the average time for warm-start ACPF with DCPF solutions as initial conditions. The rows below contain the hot-start performances as $T$, the number of samples used in CNN training, is varied. The hot-start average time is calculated by $t_{avg} = \Bar{t}_{DC} + \Bar{t}_{inf} + \Bar{t}_{AC,hot}$, where $\Bar{t}_{DC}$ is the average DCPF solution time, $\Bar{t}_{AC,hot}$ is the average hot-start ACPF solution time, and $\Bar{t}_{inf}$ is the average inference time (which is negligible compared to $\Bar{t}_{DC}$ or $\Bar{t}_{AC,hot}$).
We report the hot-start results with 1000-sample increments for $T$. We start at $T=3000$ for the \textsc{Pegase} 2869-bus case, since using a smaller $T$ produced initial conditions causing non-convergence for some load conditions. The $\Delta\mathcal{L}$ is calculated as described by Equation \ref{delL}. In particular, $\Delta\mathcal{L} = 100\%$ for the warm-start results since $\mathcal{L}_i = \mathcal{L}_f = \frac{1}{2}\Vert V_{DC} - V_{AC}\Vert_2 + \frac{1}{2}\Vert \Theta_{DC} - \Theta_{AC}\Vert_2$.

We also include the training times in the tables. These  are extremely large compared to $t_{avg}$, but they can be amortized over the usable time of the model since it is a one-time only cost. Additionally, training can be performed in parallel on clusters with multiple GPUs, which can greatly reduce the time needed. For example, prior work \cite{multi-gpu1, multi-gpu2, multi-gpu3} has shown that the training throughput (number of samples processed per second) increases quasi-linearly with the number of GPUs. The main reason they are included here is to show the trade-off between the training time and the quality of output initial conditions. 

Finally, we highlight the chosen $T$ in bold for both cases in Tables \ref{table-118} and \ref{table-2869}.

\subsection{IEEE 118-bus Results}
From Table \ref{table-118}, we can see that $T=3000$ results in a good balance of $T$ (i.e., the number of warm-start executions required to train the 1D CNN) and the quality of hot-start conditions. Even though the $T=5000$ results have both lower final $\Delta\mathcal{L}$ and solution iterations than $T=3000$ results, it has a higher $t_{avg}$ and longer training time. Compared to the warm-start results, the proposed method provides a $33.56\%$ reduction in solution time, and a $66.47\%$ reduction in the average solution iterations required, with the chosen $T$.

\begin{table}[h]
\renewcommand{\arraystretch}{1.3}
\caption{IEEE 118-bus Results}
\label{table-118}
\centering
\begin{tabular}{|p{11mm} | p{11mm} | p{10mm} | p{6mm} | p{11mm} | p{13mm}|}\hline
    \multicolumn{2}{|l|}{} & $t_{avg}$ \newline (ms) & Avg. \newline Iter. & $\Delta\mathcal{L}$ & Training \newline Time (s) \\ \cline{1-6}
    \multicolumn{2}{|l|}{Warm Start} & 1.57971 & 3.000 & 100\% & N\slash A \\ \cline{1-6}
    \multirow{5}{*}{Hot Start} & {T=1000} & 1.22120 & 1.391 & 0.2185\% & 120.96713 \\ \cline{2-6}
    & {T=2000} & 1.06184 & 1.022 & 0.0929\% & 222.45857 \\ \cline{2-6}
    & \textbf{T=3000} & \textbf{1.04953} & \textbf{1.006} &\textbf{0.05937\%} &  \textbf{326.47676} \\ \cline{2-6}
    & T=4000 & 1.05426 & 1.006 & 0.06306\% & 437.17291 \\ \cline{2-6}
    & T=5000 & 1.07008 & 1.000 & 0.03435\% & 544.79339\\ \cline{1-6}
\end{tabular}
\end{table}

\subsection{\textsc{Pegase} 2869-bus Results}
From Table \ref{table-2869}, $T=8000$ is a reasonably good choice for a larger study system, considering the rate of increase of the training time as $T$ grows, and the diminishing gain in the quality of hot-start conditions (between $T=8000$ and $T=9000$, the improvements in solution time and iterations are $0.18\%$ and $0.48\%$, respectively, but the training time increased by $1.33$ hours, or $24.65\%$).  By choosing $T=8000$, $t_{avg}$ and the average ACPF iterations required are decreased by $30.06\%$ and $49.52\%$ compared to the warm-start results, respectively. 

\begin{table}[h]
\renewcommand{\arraystretch}{1.3}
\caption{Pegase 2869-bus Results}
\label{table-2869}
\centering
\begin{tabular}{|p{11mm} | p{9mm} | p{10mm} | p{6mm} | p{11mm} | p{15mm}|}\hline
    \multicolumn{2}{|l|}{} & $t_{avg}$ \newline (ms) & Avg. \newline Iter. & $\Delta\mathcal{L}$ & Training \newline Time (s) \\ \cline{1-6}
    \multicolumn{2}{|l|}{Warm Start} & 29.71916 & 4.000 & 100\% & N\slash A\\ \cline{1-6}
    \multirow{7}{*}{Hot Start} & T=3000 & 27.99254 & 3.146 & 0.2452\% & 4098.02066 \\ \cline{2-6}
    & T=4000 & 27.11124 & 3.003 & 0.1688\% & 5450.61909 \\ \cline{2-6}
    & T=5000 & 24.46336 & 2.896 & 0.1395\% & 6475.0688 \\ \cline{2-6}
    & T=6000 & 22.86895 & 2.356 & 0.1038\% & 8952.58763 \\ \cline{2-6}
    & T=7000 & 23.40030 & 2.435 & 0.09801\% & 14539.66965\\ \cline{2-6}
    & \textbf{T=8000} & \textbf{20.78429} & \textbf{2.019} & \textbf{0.07374\%} & \textbf{19398.17254} \\\cline{2-6} 
    & T=9000 & 20.73071 & 2.000 & 0.05416\% & 24179.44558 \\\cline{1-6}
\end{tabular}
\end{table}

\section{Conclusion and Future Research Directions}
In this paper, we propose a generalizable framework to obtain better initial conditions for Newton-Raphson based ACPF using 1D CNNs. The performance of the proposed method on the IEEE 118-bus and \textsc{Pegase} 2869-bus systems show that it is capable of effectively decreasing both solution time and solution iterations. 

Although our proposed method is shown to be generalizable on both large and small systems, we acknowledge one limitation: systems with different topologies (specifically, different number of buses and/or connectivities) would require training different 1D CNNs to generate system-specific hot-start conditions. 
Consequently, we need to address the problem of long training time associated with this limitation. As we discussed in Sections \ref{s4-b} and \ref{s5}, the CNN training time can be amortized as it is a one-time cost for each system, and it can be further reduced by applying transfer learning, and by training in parallel on multiple GPUs. 

A potential research direction that builds on our proposed method 
is to incorporate the effect of ACPF solution times and/or iterations directly into the 1D CNN training stage, e.g., through a similar meta-optimization step as discussed in \cite{opf-meta}. We can also include different contingency scenarios in the dataset generation step, and train 1D CNNs to produce initial conditions to perform contingency analysis more efficiently compared to the DCPF model. 
Finally, we could also try out different CNN architectures, e.g., deeper architectures such as the fully convolutioonal ResNet in \cite{fcresnet}, although they will require significantly longer training time and a much larger and diverse dataset to be properly trained. \\

\appendix\label{appendix}
\begin{center}
    \textsc{\small CNN Overview} \\
\end{center}

We give a brief overview of CNNs and the terms used in previous sections based mostly on \cite{Goodfellow2015DL}. The most important building block of a CNN is the convolutional layer, which contains the ``filter" or ``kernel" and the output called the ``feature map." These are controlled by adjustable hyperparameters such as kernel size (height, width, channels), stride, and zero padding size. In our 1D CNN, kernel size refers to the dimension of the kernel along the height dimension. The kernel contains the parameters of the CNN to be updated based on the loss computed by a loss function. The kernel values can be initialized randomly, or via a scheme such as the Xavier Initializtion \cite{xavier}. The operation of the kernel on the input is the 1D convolution commonly seen in many areas of engineering. The ``stride" refers to the step size of the moving kernel, which in our case is 1. Our chosen architecture ends with a fully connected layer, which produces a desired number of values each resembling the target values during training, based on the outputs of the last convolutional layer. The parameters in this fully connected layer are the edge weights in a complete bipartite graph between the set of reshaped output of the last convolutional layer and the set of final outputs. 

The number of ``channels" originates from the RGB channels of images. In our case, the input channel of size 4 represents the four signals in each bus that we consider (real and reactive power, voltage magnitude and phase). The kernel channel sizes have no explicit meaning, but can be roughly thought as filters that extract different hidden features in the data.

\begin{figure}[h]
    \centering
    \includegraphics[trim={2.75cm 4.5cm 2cm 2cm},clip,scale=0.325,center]{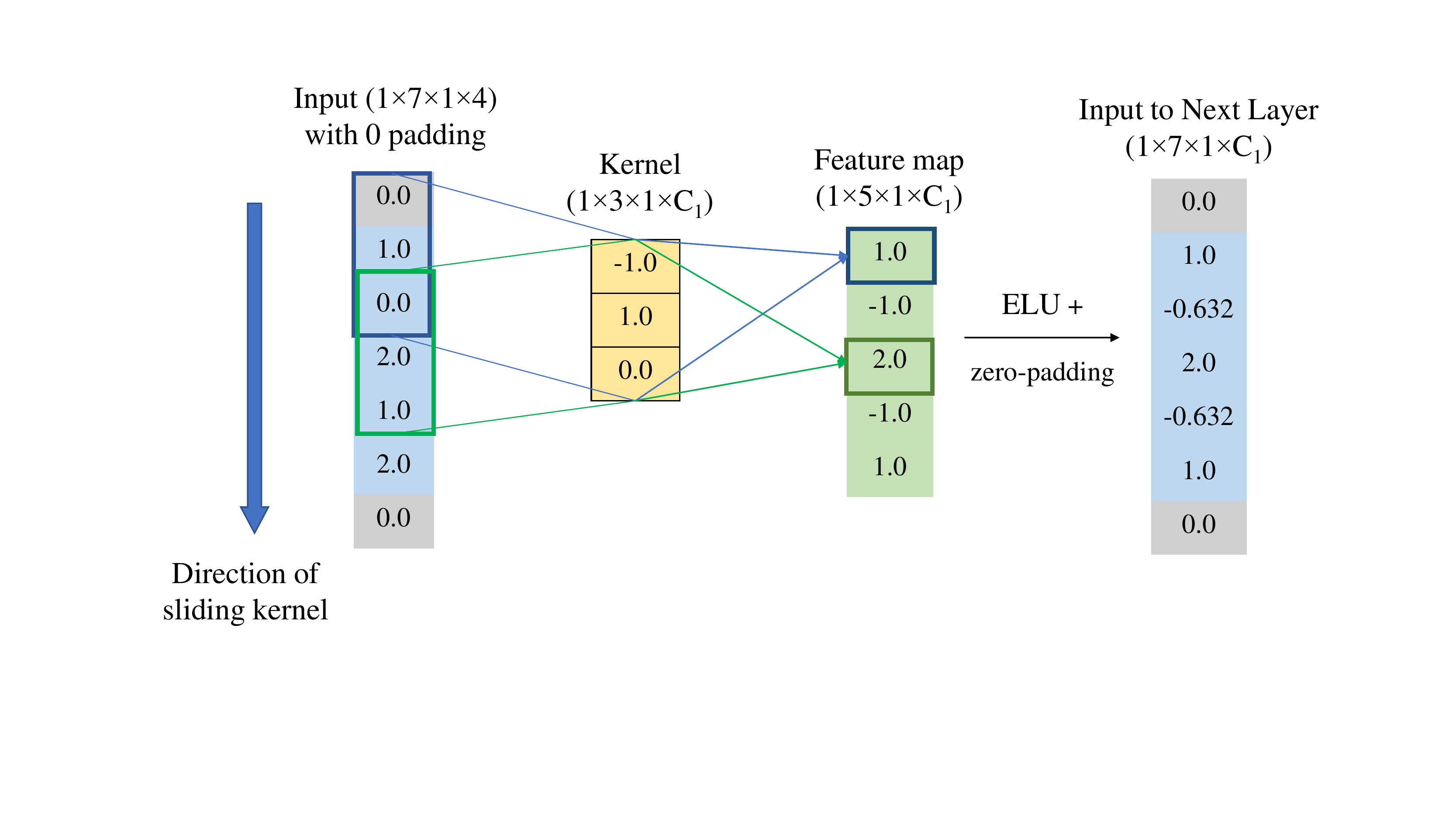}
    \caption{Simple 1D CNN example, 5-bus case with arbitrary values and a kernel size of 3. ($C_1 = 8$ from Figure \ref{model}.)}
    \label{cnnex}
\end{figure}

Controlling the subsequent feature maps dimensions, as we mentioned before, is helpful since we can avoid upsampling layers by keeping the input and output dimension the same. Figure \ref{cnnex} demonstrates a simple example of a single convolutional layer, and only the first channel of each component is shown. To visualize the convolution operation on all four channels, refer to the demonstration in \cite{stanford_conv} in the context of the dataset format described in Section \ref{s3-c}. 

{}

Finally, training of CNNs are now almost always done on hardware accelerators designed for large parallel computations, such as GPUs and TPUs. Training samples are usually fed into the CNN in mini-batches with size between 1 and $N$, and the errors at the output are back-propagated into the CNN for parameter updates. Separating training samples into mini-batches is preferred since training with a single sample at a time results in more frequent, thus computationally more expensive (and less accurate) parameter updates. On the other hand, training by feeding in the full dataset can require large amounts of memory to store all training samples, and can be prone to erroneous early convergence due to local minima. During training, ``epochs" represent the number of times that the CNN processes the entire training set, which typically range from a few dozen to a few hundred. \\*

\bibliographystyle{IEEEtran}
\bibliography{ref}

\end{document}